# Measurement of the Zak phase of photonic bands through the interface states of metasurface/photonic crystal

Qiang Wang[1,*], Meng Xiao[2*], Hui Liu[1,+], Shining Zhu[1] and C.T. Chan[2]

[1]*National Laboratory of Solid State Microstructures & School of Physics, Collaborative Innovation Center of Advanced Microstructures, Nanjing University, Nanjing, 210093, China*

[2]*Department of Physics and Institute for Advanced Study, the Hong Kong University of Science and Technology, Clear Water Bay, Kowloon, Hong Kong*

[*]These authors contributed equally to this work.
[+] Correspondence address: liuhui@nju.edu.cn

**Zak phase labels the topological property of one-dimensional Bloch Bands. Here we propose a scheme and experimentally measure the Zak phase in a photonic system. The Zak phase of a bulk band of is related to the topological properties of the two band gaps sandwiching this band, which in turn can be inferred from the existence or absence of an interface state. Using reflection spectrum measurement, we determined the existence of interface states in the gaps, and then obtained the Zak phases. The knowledge of Zak phases can also help us predict the existence of interface states between a metasurface and a photonic crystal. By manipulating the property of the metasurface, we can further tune the excitation frequency and the polarization of the interface state.**

Topological invariant plays a more and more important role in modern physics with the discovery of new materials such as topological insulators [1,2]. The concept of momentum space topology has also been extended to various photonic systems [3-16] to realize interesting applications. In a two-dimensional system, the topological invariant is characterized by the first Chern number [17,18], which is proportional to the Berry phase [19] enclosing the first Brillouin zone. In a one-dimensional (1D) system [20,21], the topological invariant can be characterized by the Zak phase [22], a special kind of Berry phase defined along a 1D bulk band.

One of the major challenge about the topological invariant is how to measure it, which is quite difficult as it is an abstract concept and hence hard to be observed directly. In recent years, topological invariants have been measured in cold atom systems [23-25] and an acoustic system [26]. A few methods have also been proposed to measure topological invariants in electromagnetic wave systems [27-33]. Recently, Xiao *et.al*, theoretically investigated the relationship between the Zak phases and the surface impendence in 1D PCs [34]. Inspired by this work, here we propose a method and experimentally measure the Zak phases through interface states.

The schematic diagram of our system is illustrated in Fig. 1(a), which is composed of a metal film (or metasurface) and a PC. Detail structural is given in Fig. 1(b). A unit cell composes of A/B/A layers as marked by the red dashed lines in Fig. 1(b). Here, the layer A is HfO$_2$ with thickness $t_A = 177.6 nm$ and the layer B is SiO$_2$ with thickness $t_B = 579.8 nm$. The unit length is $\Lambda = 2t_A + t_B = 935 nm$. We calculate its band dispersion as shown in Fig. 2(a) which covers the range $1.1 < \Lambda/\lambda < 1.9$, including two bands (3rd and 4th band counting from low frequency to high frequency) and three gaps (2nd, 3rd and 4th gap). The refractive indexes of HfO2 and SiO2 used are from experimental measurements [35]. In the working frequency range, the dispersion is negligible, and the refractive indexes of HfO2 and SiO2 are around 2 and 1.46, respectively. Zak phase is defined as [22]:

$$\theta_n^{Zak} = \int_{-\pi/\Lambda}^{\pi/\Lambda} \left[ i \int_{\text{unit cell}} dz \varepsilon(z) u_{n,q}^*(z) \partial_q u_{n,q}(z) \right] dq, \qquad (1)$$

where $\varepsilon(z)$ is the relative permittivity, $u_{n,q}$ is the normalized periodic part of the electric field Bloch eigenfunction of a state on the $n^{th}$ band with wave vector $q$. As the value of Zak phase depends on the choice of origin [36], here we choose it to be the interface of the two layer A, which is also the boundary of the unit cell. The values of Zak phases of band 3 and 4 are calculated and labled in green in Fig. 2(a), which are 0 for band 3 and $\pi$ for band 4.

Besides Eq. (1), we can also determine the Zak phase using the symmetry of the band edge states. As the unit cell possesses mirror symmetry, the band edge states are either even or odd with respect to mirror plane. Here, the symmetry types of band edge states are defined using the symmetry of eigen electric field (see supplementary document). The symmetry of the six band edge states are labeled with red (symmetric) and blue (antisymmetric) circles respectively in Fig. 2(a). If the two band edge states of a same band have the same symmetry, the Zak phase of this band is 0, otherwise, the Zak phase is $\pi$ [22,37]. However, the determination of the symmetry types of band edge states is quite difficult to implement experimentally in optical frequency.

According to the surface bulk correspondence [34], the Zak phase is related to the signs of the reflection phases, i.e.,

$$\exp(i\theta_n^{Zak}) = -sgn(\varphi_n) / sgn(\varphi_{n-1}), \qquad (2)$$

where $\theta_n^{Zak}$ is the Zak phase of the $n^{th}$ bulk band, $\varphi_n$ and $\varphi_{n-1}$ are the reflection phases of the $(n-1)^{th}$ gap (gap blow the $n^{th}$ bulk band) and $n^{th}$ gap (gap above the $n^{th}$ bulk band), respectively. Measuring the value of the reflection phase inside a band gap may need complex interference experimental setup [38]. However, Eq. (2) shows that we only need the sign, which greatly simplifies the measurement.

As the two band edge states sandwiching a gap must possess different symmetry types as required by the orthogonality, there are only two combinations: the first combination is symmetric lower band edge state and antisymmetric higher band edge state, which is denoted as S-A, for instance, gap 3 and 4. The other is antisymmetric lower band edge state and symmetric higher band edge state (A-S), for example, gap 2. The difference between these two combinations is manifested in their reflection phase range [34]. For S-A type gaps, the reflection phase belongs to $(0,\pi)$; and that of A-S type belongs to $(-\pi,0)$, similar to metal of which the sign of reflection phase is also negative. Note that the reflection phase inside a band gap region of a 1D PC does not change sign [34]. We numerically calculate the reflection phase ($\varphi_{PC}$) as shown in Fig. 2(b) in gray. The reflection phases inside gap regions are highlighted in red. The reflection phase of gaps 3 and 4 are positive and highlighted with yellow color. While that of gap 2 is negative and is highlighted in blue. In experiment, we fabricated one PC sample according to the design in Fig. 1(b) with the same parameters as used in Fig. 2. The sample includes twelve periods (A/B/A). We measured the reflection spectrum of the sample and the result is shown as the black curve in Fig. 3(a), which matches well with numerical simulation shown as the black curve in Fig. 3(b).

Tamm plasmon state [39,40] can be constructed in the system if we deposit a silver film on the PC. In experiment, the thickness of the silver film is 70nm. The negative value of its reflection phase ($-\varphi_{Ag}$) is numerically calculated and shown in Fig. 2(b) with purple lines. The existence condition of an interface state is governed by the equation $\varphi_{PC} + \varphi_{Ag} = 0$. And hence if there is an interface state inside a gap region, we know the $\text{sgn}(\varphi_{PC})$ inside this gap must be positive. Otherwise, $\text{sgn}(\varphi_{PC})$ is negative. The crossing points in band gaps 3 and 4 satisfy $\varphi_{PC} + \varphi_{Ag} = 0$, which indicate the existence of an interface state. As a comparison, there is no crossing point in band gap 2, which means no interface state inside gap 2. According to the discussion above, Eq. (2) can be rewritten as:

$$\theta_n^{Zak} = \begin{cases} 0 & V_{n-1} \neq V_n \\ \pi & V_{n-1} = V_n \end{cases}, \quad (3)$$

where $V_{n-1}$ and $V_n$ indicate whether there exists an interface state or not in the lower and upper gaps of the $n^{th}$ band respectively. If the $n^{th}$ gap possesses interface states, $V_n = 1$, otherwise $V_n = 0$. Eq. (3) means, if the two adjacent band gaps possess or do not possess interface states simultaneously, the Zak phase of that band takes value $\pi$, otherwise 0. Experimentally, the existence of an interface state can be seen from the reflection spectrum which is presented as a dip inside the gap region. In Fig. 3(a), the red curve shows the reflection spectrum measured with normal incidence. The two sharp dips in gap 3 and gap 4 indicate the existence of interface states ($V_{3,4} = 1$), while there is no interface state in band gap 2 ($V_2 = 0$), and the corresponding results are summarized in Table. I, where yellow (blue) means there exists (does not exist) an interface state inside this gap. Besides little global shift, which we think stems from the difference between the refractive index of materials in the experiments and those in the theoretical model, all the results match well with numerical simulations in Fig. 3(b). According to Eq. (3), we can obtain the values of the Zak phase of band 3 as $\theta_3^{Zak} = 0$, and band 4 as $\theta_4^{Zak} = \pi$.

| Gap 2 | Zak phase of Band 3 | Gap 3 | Zak phase of Band 4 | Gap 4 |
|---|---|---|---|---|
| A-S $\{-\pi, 0\}$ | 0 | S-A $\{0, \pi\}$ | $\pi$ | S-A $\{0, \pi\}$ |

TABLE I The symmetry of band edge states, the reflection phase range of band gaps and the Zak phases of bulk bands. The yellow (blue) region represents the existence (the absence) of an interface state inside this gap.

In the above investigation, the interface state is localized at the boundary between a silver film and a PC, whose frequency is determined by the structural parameters of the PC and the sliver film. Different from the zero-energy edge states protected by the chiral symmetry [41], the interface state here is protected by the inherent mirror symmetry of the PC. To predict the existence of an interface state, we do not need to identify any topological transitional points [42,43]; instead we only need the topological properties of the gaps on both sides of the interface to be different, which are related to the summation of the Zak phases of all the bulk bands below this gap [34]. We note that in crystalline topological insulators [44], the edge termination must be compatible with the symmetry used to define the bulk topology in order to guarantee bulk-edge correspondence. In our system, the edge termination must be at mirror planes in order for the bulk-edge correspondence to predict the existence of edge modes.

In applications, it is highly desirable that the working frequency can be tuned without changing the PC structure, and this can be achieved by replacing the metal film with a metasurface. Metasurfaces, which are composed of optical metamaterials in a reduced dimensionality, are applied to many novel interesting applications [45-62]. If we replace the sliver film with metasurfaces, the interface state in the metasurface/PC configuration can be manipulated through engineering the reflection phase of the metasurface. Our metasurface possesses strong anisotropy and is designed to be composed of a subwavelength grating and a thin $Si_3N_4$ film. With this metasurface, we are able to manipulate both the excitation frequency and the polarization of the interface states.

The structure of our metasufce/PC configuration is showed schematically in Fig. 1(b). The structure parameters of PC are $t_A = 97.5 nm$ and $t_B = 267.0 nm$. The metasurface is composed of subwavelength silver grating and a thin $Si_3N_4$ film. The slits in the grating is etched from a 50nm-thick sliver film by Focused Ion Beam (FIB), whose width can be precisely controlled in the process. Between the PC and the grating, we insert a 55nm-thick (much less than the working wavelength at about 1.5 um) $Si_3N_4$ film by Plasma Enhanced Chemical Vapor Deposition (PECVD), which protects the surface of PC from being damaged by FIB. The nano slits are periodical along the X direction with thickness $h = 50nm$ and period $P = 300nm$. The width of a single slit is $d$, which is a tunable parameter in experiments. We fabricate five samples with different slit widths, $d$ =0nm (equivalent to a sliver film), 50nm, 100nm, 150nm and 200nm. The calculated negative value of reflection phases of the metasurfaces with the Y polarization (electric along the Y direction) and the X polarization (electric along the X direction) are shown in Fig. 4(a) and (b), respectively. The black, red, green, blue and cyan lines represent the silt width $d$=0, 50, 100, 150 and 200nm, respectively. For reference, the calculated reflection phase and the measured reflection spectrum of our PC are also shown in Fig. 4 and Fig. 5 as orange lines. Comparing Figs. 4(a) with (b), the reflection phases of the Y and the X polarization have the opposite trends with the increasing of $d$ (denoted by arrows in Fig. 4). Due to the reflection phase change of the metasurface, the crossing points inside the gap region shift accordingly, and hence interface states with different polarizations will shift in different directions. For the Y polarization, the curve of metasurface shifts downward with increasing $d$ (denoted by blue arrow), and always has a crossing point with the curve of the PC in the gap region. However, for the X polarization, the curve of metasurface shifts upward (denoted by red arrow), and when d >100nm, there is no crossing point and hence no interface state. The different responses of different polarizations lie in the fact that the metasurface is highly anisotropic. It behaves like a metal slab for the Y polarization and a dielectric slab for the X polarization. For the Y polarization, the existence of interface state is protected and we can only shift the frequency of the interface state inside the band gap.

The above results can also be seen from the reflection spectrum, which is measured by Fourier Transformation Infrared Spectroscopy (FTIR) with normal incidence . The measured data is shown in Fig. 5(a) for the Y polarization and Fig. 5(b) for the X polarization, respectively. The sharp dips inside the band gap on the reflection spectrums indicate the existences of interface states. In Fig. 5(a), the frequency of the dip is red-shifted with increasing values of *d*. Consistent with the absence of crossing inside the gap region in Fig. 4(b) for the *d*=100, 150 and 200nm cases, we cannot find reflection dips for those three cases. The dip of the *d*=50nm case gets a little mixed up with the bulk band Fabry–Pérot crinkle as it is near the band edge and the PC is finite in the experiment. As a comparison, we also present the numerical results in Figs. 5(c) and (d) for the Y and the X polarization, respectively. The numerical and experimental results match well with each other. Our results demonstrated the polarization selection of the interface states using the metasurface/PC configurations.

Due to the large field intensity enhancement, the interface state can be used in those applications which require large local fields, such as controlling the spontaneous optical emission from quantum dots [63], realizing single photon source [64] and demonstrating nanolasers [65,66]. In addition, the meta-superstructure at the photonic crystal surface can be designed to manipulate the interface state for applications such as optical switches [67] and sensors [68].

In a summary, we implemented an experiment to measure the Zak phases in a layered optical system. Instead of probing the reflection phase or the symmetry of band edge states, we determined the Zak phase through the interface state, which is indicated in the reflection spectrum. Although our focus here is in one-dimensional system, this method can be extended to 2D or 3D systems.

Besides, we showed the manipulation of interface states including both the excitation frequency and the polarization by introducing metasurfaces. The flexible control of the interface states at the boundary of meatsurface/PC is an area that deserves more investigation and may lead to new applications.

This work was financially supported by the National Natural Science Foundation of China (No. 11321063, 61425018 and 11374151), the National Key Projects for Basic Researches of China (No. 2012CB933501 and 2012CB921500), Hong Kong Research Grant Council grant AOE/P-02/12 and the SRFDP/RGC ERG Joint Research Scheme M-HKUST601/12, the Doctoral Program of Higher Education (20120091140005), and Dengfeng Project B of Nanjing University.

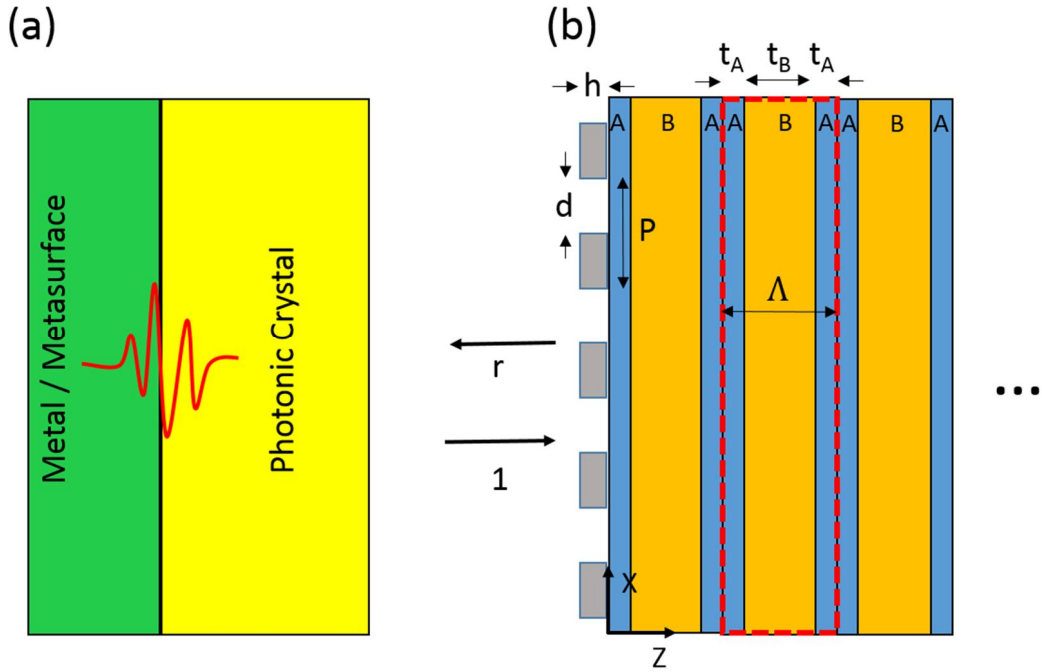

**Fig. 1** (a) Sketch of an interface between a metal film (or metasurface) and a photonic crystal which supports interface states. (b) Sketch of the metasurface/photonic crystal system. The metasurface consists of nano slits etched on a sliver film with width $d$, period $P$ and thickness $h$. The period of the photonic crystal is $\Lambda$, with A-B-A as its unit cell as marked by the red dashed lines. Here, layer A is $HfO_2$ and layer B is $SiO_2$.

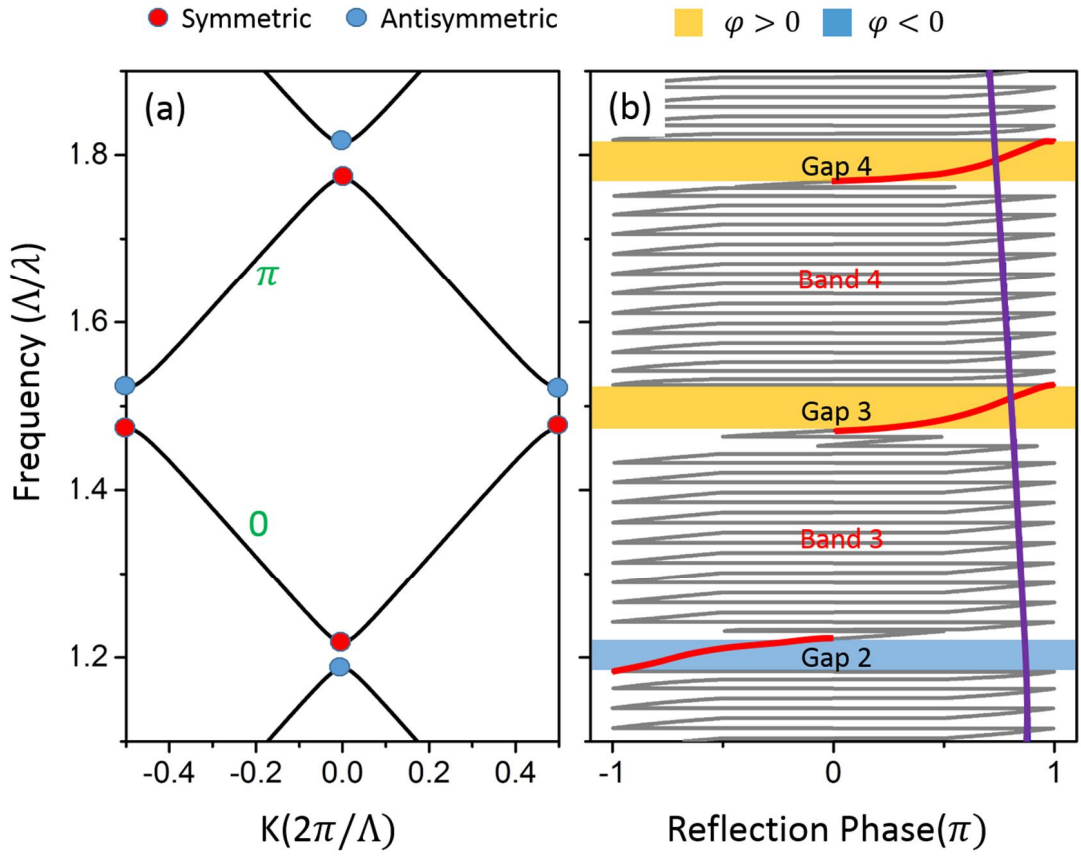

**Fig. 2** (a) The band structure of the photonic crystal (PC) shown in Fig. 1(b), with $t_A = 177.6nm$ and $t_B = 579.8nm$. Zak phases are labeled in green. (b) The gray curves represent the reflection phase ($\varphi_{PC}$) of the 1D PC consisting of 12 periods. The reflection phase inside the gap regions are highlighted in red. The yellow (blue) region represents gap with reflection phase $\varphi_{PC} > 0$ ($\varphi_{PC} < 0$). The purple line shows the negative value of reflection phase ($-\varphi_{Ag}$) of a 70nm-thick silver film.

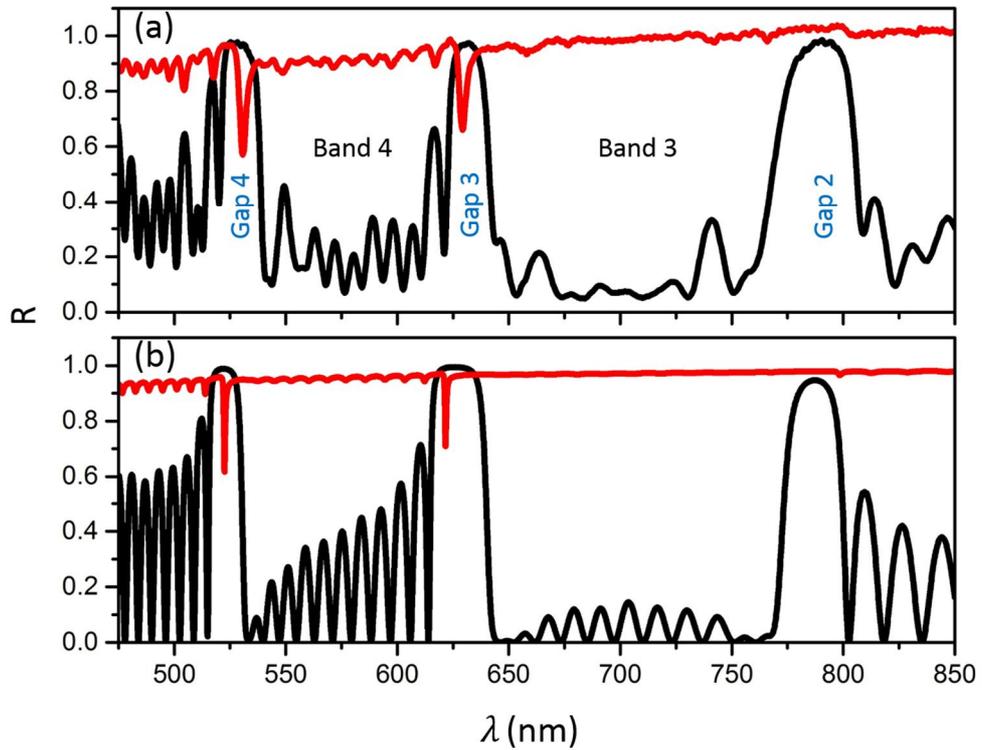

**Fig. 3** Experimental (a) and numerical (b) reflection spectrum of the PC (black line) and the silver film/PC (red line). The sharp dips of the red curve inside the gap frequency regions of the PC represent interface states between the silver film and the PC.

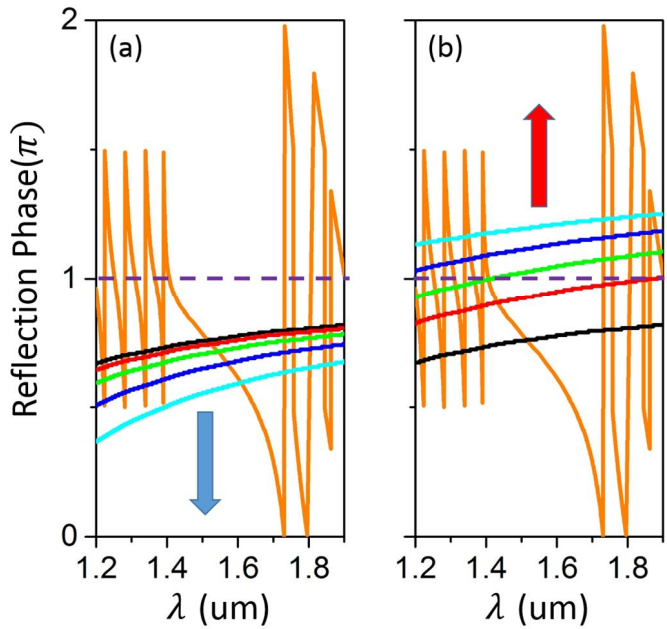

**Fig. 4** The calculated negative value of reflection phase of the metasurfaces (including the 55nm thick Si$_3$N$_4$ film) with the Y polarization (a) and the X polarization (b). The black, red, green, blue and cyan lines represent silt widths *d*=0 (equivalent to sliver film), 50, 100, 150 and 200nm respectively. The orange line shows the reflection phase of the PC with $t_A = 97.5nm$ and $t_B = 267.0nm$. The number of periods of the photonic crystal is 16.

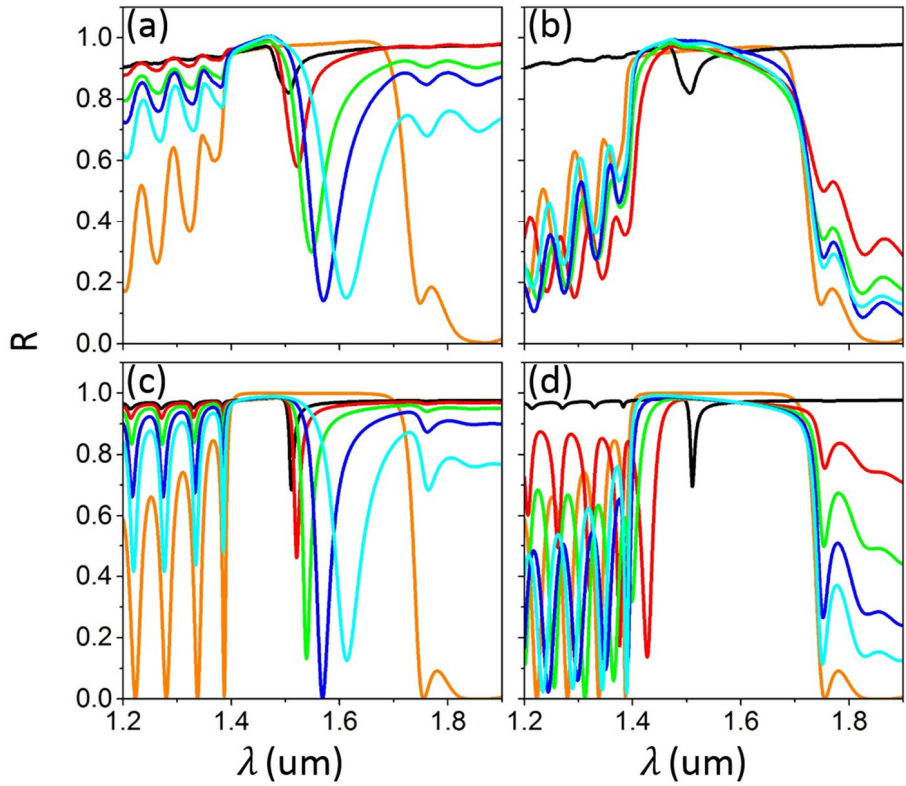

**Fig. 5** Reflection spectrum of the metasurface/PC shown in Fig. 1(b), with the Y polarization (a, c) and the X polarization (b, d), where a, b (c, d) are experimental (numerical) results. The black, red, green, blue and cyan lines represent silt widths *d*=0 (equivalent to sliver film), 50, 100, 150 and 200nm respectively. The orange line represents the reflection spectrum of the PC.

# Supplementary Material

## of

## Measurement of the Zak phase of photonic bands through the interface states of metasurface/photonic crystal

Qiang Wang[1,*], Meng Xiao[2,3,*], Hui Liu[1,+], Fan Zhong[1], Shining Zhu[1] and C. T. Chan[2,3,+]

[1]National Laboratory of Solid State Microstructures & School of Physics, Collaborative Innovation Center of Advanced Microstructures, Nanjing University, Nanjing, 210093, China

[2]Department of Physics, Hong Kong University of Science and Technology, Clear Water Bay, Kowloon, Hong Kong

[3]Institute for Advanced Study, Hong Kong University of Science and Technology, Clear Water Bay, Kowloon, Hong Kong

1. **Symmetry of band edge states in PC1**

For PC1, its structural parameters are $t_A = 177.6 nm$, $t_B = 579.8 nm$. Its dispersion as shown in Fig. S1 is obtained from the following equation [1]:

$$\cos(q\Lambda) = \cos(2k_A t_A)\cos(k_B t_B) - \frac{1}{2}\left(\frac{n_B}{n_A} + \frac{n_A}{n_B}\right)\sin(2k_A t_A)\sin(k_B t_B), \quad (S1)$$

where $k_{A,B} = \omega n_{A,B}/c$, $n_A$ and $n_B$ are the refractive index of layer A and B, respectively. The symmetry type of each band edge state M, N, O, P, Q, R in Fig. S1 can be determined with the transfer-matrix method in Ref. [1]. In our experiment, we consider normal incidence and assume the electric field is along the Y direction (as defined in Fig. 1(b) of the main text). The electric field in layer A is

$$E_y(z) = t_{12}\exp(ik_A z) + (\exp(iq\Lambda) - t_{11})\exp(-ik_A z), \quad (S2)$$

with

$$t_{11} = \exp(i\,2k_A t_A)\left[\cos(k_B t_B) + \frac{i}{2}\left(\frac{n_B}{n_A} + \frac{n_A}{n_B}\right)\sin(k_B t_B)\right], \quad (S3)$$

$$t_{12} = \exp(-i\,2k_A t_A)\left[\frac{i}{2}\left(\frac{n_B}{n_A} - \frac{n_A}{n_B}\right)\sin(k_B t_B)\right], \quad \text{(S4)}$$

The electric field in layer B is

$$E_y(z) = s_{11}\exp(ik_B z) + s_{12}\exp(-ik_B z), \quad \text{(S5)}$$

with transfer matrix equation

$$\begin{pmatrix} \exp(i\,2k_B t_A) & \exp(-i\,2k_B t_A) \\ \exp(i\,2k_B t_A) & -\exp(-i\,2k_B t_A) \end{pmatrix}\begin{pmatrix} s_{11} \\ s_{12} \end{pmatrix}$$
$$= \begin{pmatrix} \exp(i\,2k_A t_A) & \exp(-i\,2k_A t_A) \\ \dfrac{n_a}{n_b}\exp(i\,2k_A t_A) & -\dfrac{n_a}{n_b}\exp(-i\,2k_A t_A) \end{pmatrix}\begin{pmatrix} t_{12} \\ \exp(iq\Lambda) - t_{11} \end{pmatrix}, \quad \text{(S6)}$$

With Eqs. (S2 – S6), we plot the absolute values of of eigen electric fields ($|E_y|$) for the six band edge states (labled with M, N, O, P, Q and R as shown in Fig. S1) in Fig. S2. The red dashed line marks the mirror center as well as the origin in our system. The eigenfield of an antisymmetry state satisfies $E_y(z) = -E_y(-z)$, and hence $|E_y(z)| = 0$, so M, P and R belong to antisymmetric states; The eigenfield of a symmetry state satisfies $E_y(z) = E_y(-z)$, and hence $|H_x(z)| = 0$ and $|E_y(z)|$ reaches its local maximum at $z = 0$, so N, O and Q belong to symmetric states.

2. **Band structure and symmetry of band edge states in PC2**

For PC2, its strutural parameters are $t_A = 97.5nm$ and $t_B = 267.0nm$. The band dispersion of PC2 around its first gap is shown in Fig. S3(a). The absolute values of the eigenfields $|E_y|$ of the two band edge states (M and N) sandwiching the first gap are shown in Fig. S3(b) and (c), respectively. According to the discussion in the main text, we know that the values of the reflection phase inside the band gap belong to $\{0, \pi\}$.

## 3. Reflection spectrum of metasurfaces

Besides the reflection phase, the reflection spectrum also changes with the slit width of the metasurface. The numerically obtained reflection spectrums are shown in Fig. S4 for the normal incident wave with (a) Y polarization and (b) X polarization. The black, red, green, blue and cyan lines represent the widths of silts with $d$=0 (equivalent to sliver film), 50, 100, 150 and 200nm respectively. We also show the reflection intensities of the PC2 as orange lines as reference. Comparing Fig. 4(a) and (b), the reflection intensities show different responses to the Y and X polarization with the change of structural parameters in metasurfaces.

## 4. Interface state of metasurface/PC2

To show that the intensity of the interface state is well localized in the interfical region as shown in Fig. S5(a), we numerically calculate the intensity distribution with slits width $d = 150nm$ for the normal incident wave with (b) Y polarization and (c) X polarization. The wavelength (1.570um) here is at the excitation wavelength of the interface state for Y polarization. The black dashed line marks the interface between the metasurface and PC2 (in the right side). The orange dashed line frames shows the position of the sliver strips and the boundary of the two periods is labeled with white dashed line. The red region in Fig. S5(a) shows about 20-fold enhancement of local field intensity at the excitation wavelength of the interface state, while there is no enhancement with X polarization in Fig. S5(b). In addition, we also calculated the intensity distribution at different wavelength in the gap region as shown in Fig. S6 with Y polarization (a) and X polarization (b). The horizotal axis labels the position of the center of the slit as shown in Fig. S5 with white dashed lines. And vertical axis labels wavelength in the band gap. It is clear from Fig. S6 (a) with (b) that the interface state has dependence on the polarization of the incident wave.

5. **The excitation wavelength of the interface states with the different metasurfaces**

To distinctly show the trends of the interface states with the changing of metasurfaces, we extract the excitation wavelength with different slit width *d* from the measured reflection spectrum for Y polarization as shown in Fig. 5(a) in main text, and show in Fig. S7 with red stars. The numerical results are also shown with black crosses for comparison. Numerical and experimental results match very well.

**Reference:**

[1]   A. Yariv and P. Yeh, Optical waves in crystals: propagation and control of laser radiation (Wiley, New York, 2003)

**Figures:**

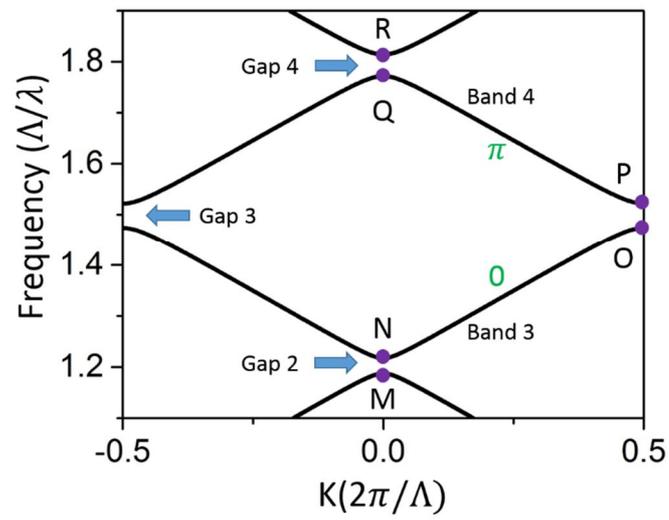

Fig. S1 The band dispersion of PC1 with structural parameters $t_A = 177.6nm$, and $t_B = 579.8nm$. The six purple circles mark the positions of the band edge states. The Zak phases of the 3$^{rd}$ and the 4$^{th}$ band are given as in green numbers.

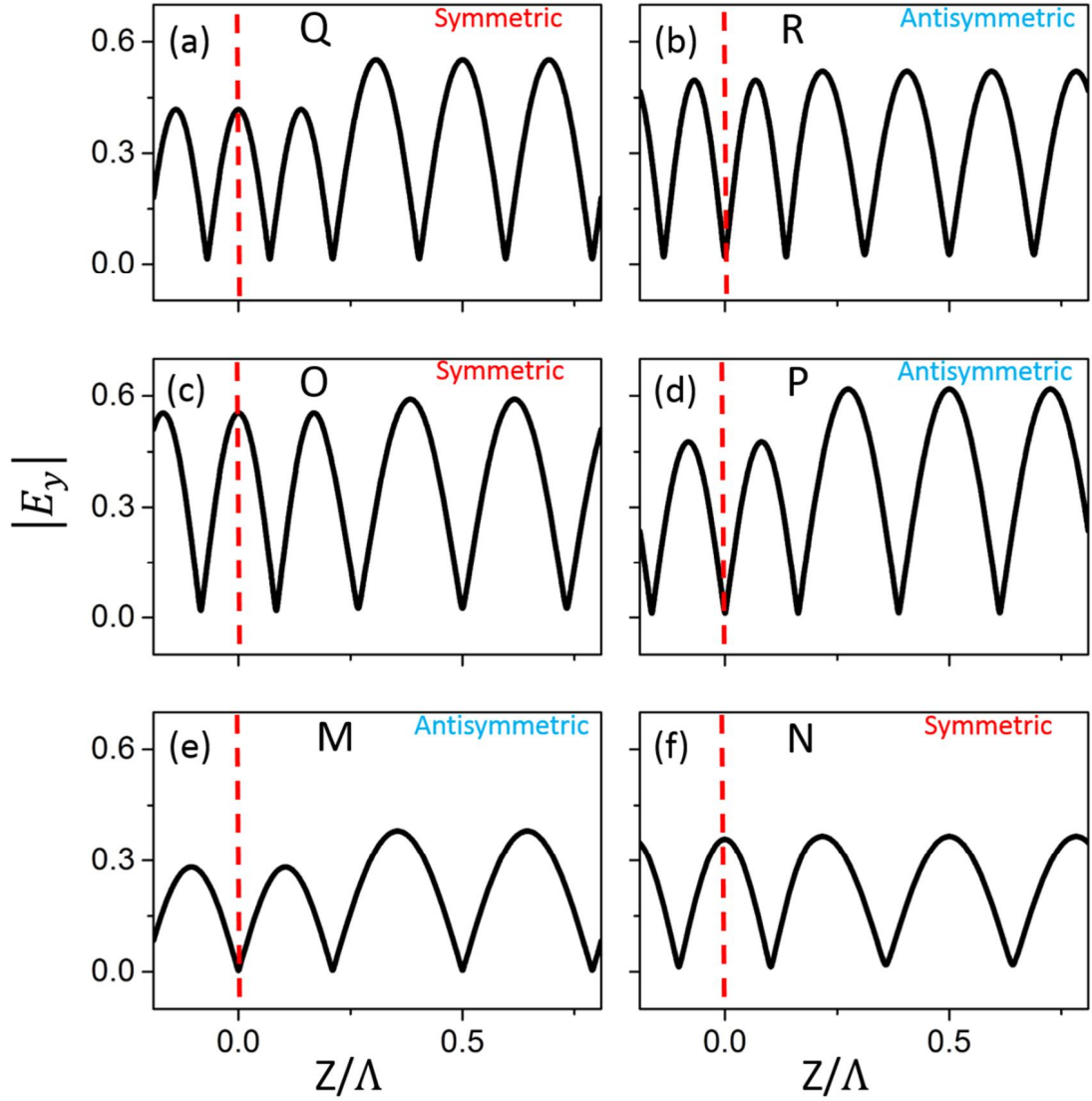

Fig. S2 The amplitude of the electric fields of the band edge states as a function of z. (a)-(f) are for the six band edge states, R, Q, P, O, N, M, which are labeled in Fig.S1. The red dashed lines indicate the location of interface between two A layers.

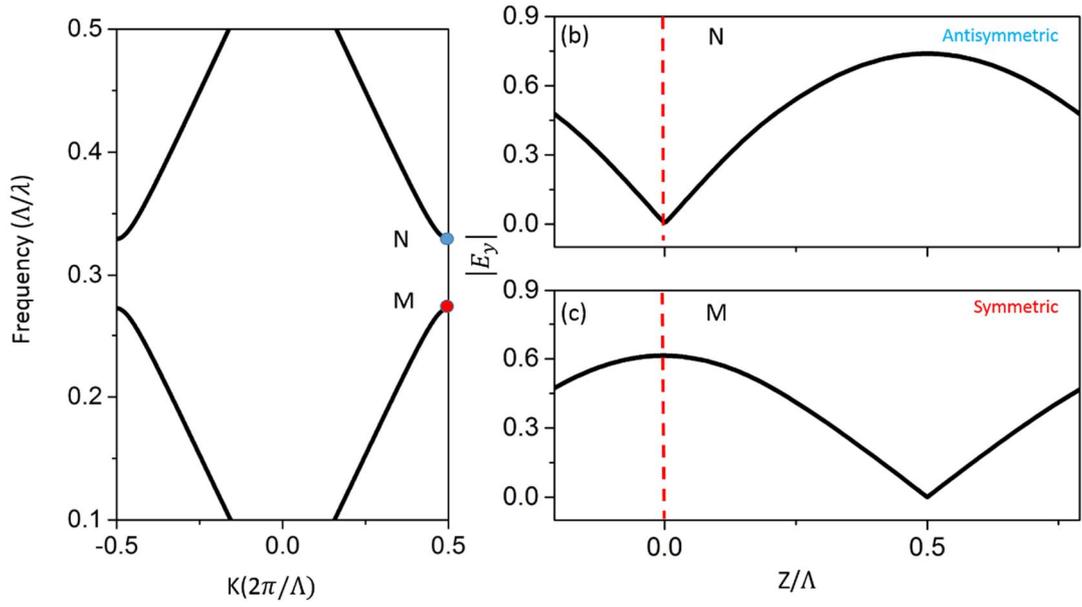

Fig. S3 (a) The band dispersion of the PC2 with layer-A thickness $t_A = 97.5nm$, and layer-B thickness $t_B = 267.0nm$. M, N label the two band edge states of the band gap we are interested in. (b) and (c) are the $|E_y(z)|$ of band edge states N and M. The red dashed lines indicate the location of interface between two A layers.

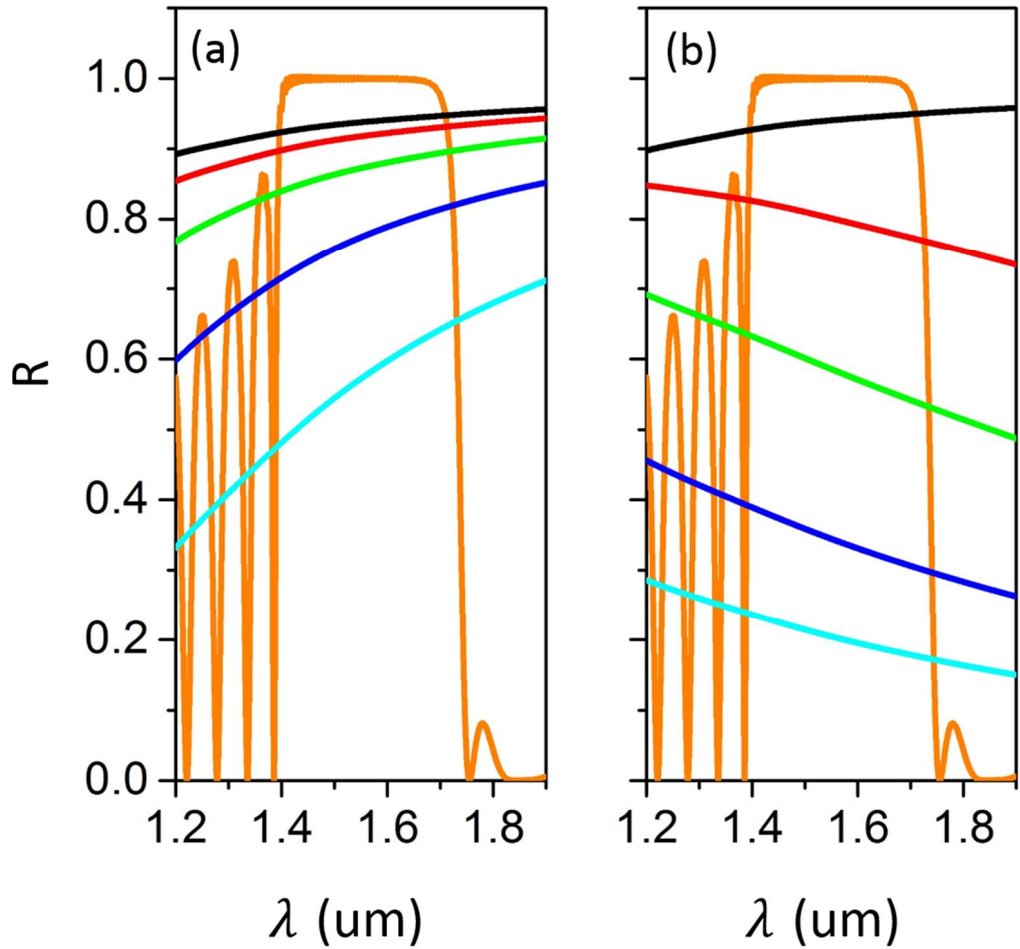

Fig. S4 The calculated reflection spectrum of the metasurfaces (including the 55nm thick $Si_3N_4$ film) of a normal incident wave with (a) Y polarization and (b) X polarization. The black, red, green, blue, and cyan lines represent the widths of silts with *d*=0 (equivalent to a silver film), 50, 100, 150 and 200nm. The orange line shows the reflection spectrum of the PC2 for reference.

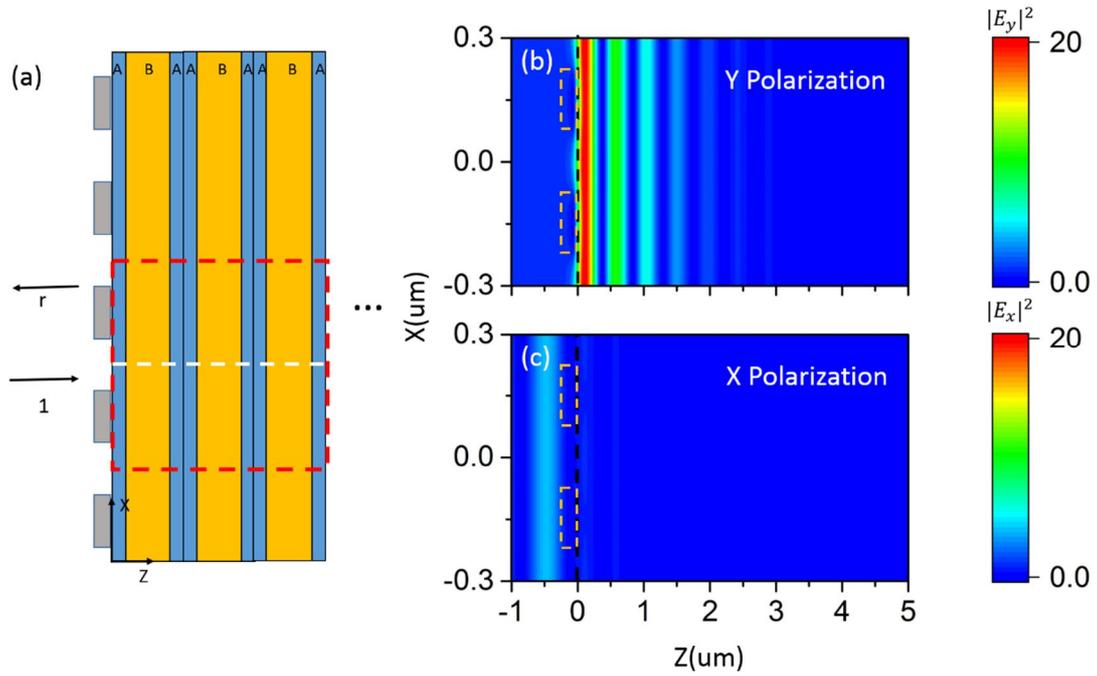

Fig. S5 Schematic diagram of the interface state is showed in (a). The distribution of $|E|^2$ at the interface of metasurface/PC2 for the wavelength 1.570um with (b) Y polarization and (c) X polarization. Two periods of the metasurface marked with red dashed lines in (a) is shown in the figures with slits width *d*=150nm. The boundary of between two unit cells is marked with white dashed line. The black dashed line marks the left boundary of PC2, and the orange dashed lines show the position of the silver strips which are the metallic component of the metasurface. The wavelength here is at the excitation wavelength of the interface state for Y polarization (b), while there is no interface state for X polarization (c).

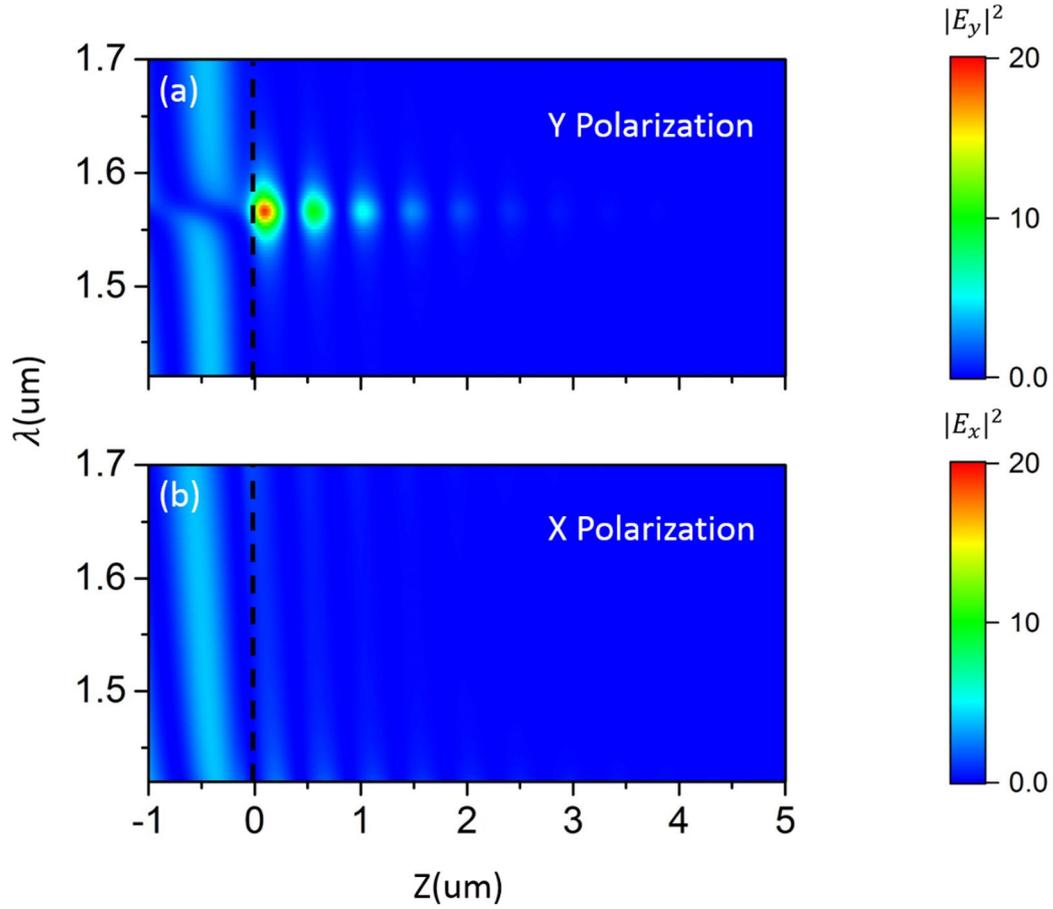

Fig. S6 The distribution of $|E|^2$ along the white dashed line in Fig. S5(a) for the incident wavelength 1.450~1.70um with (a) Y polarization and (b) X polarization. The black dashed lines mark the interface between the metasurface and PC2. The position where field is calculated is the center of two adjacent silts as shown in Fig. S5(a) with white dashed lines. Y axis denotes the incident wavelength.

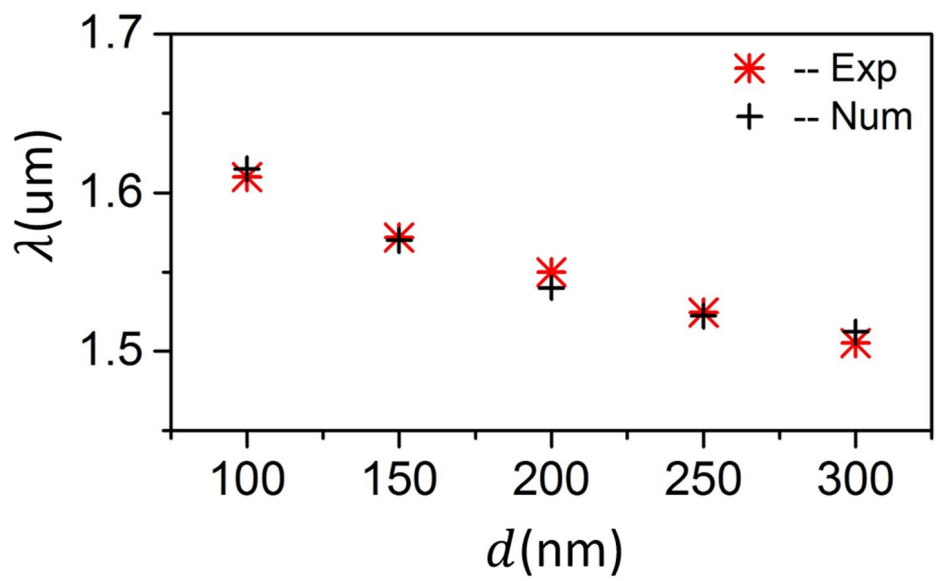

Fig. S7 The excitation wavelength of the interface states for Y polarization with the slit width $d$. The red stars represent the excitation wavelength of the measured interface states, which are measured in experiments, and the black crosses mark the numerical results.